\documentclass[12pt,pra,a4paper,superscriptaddress,onecolumn]{revtex4}
\usepackage[english]{babel}
\usepackage[T1]{fontenc}
\usepackage[utf8]{inputenc}
\usepackage{graphicx,epstopdf}
\usepackage{amssymb}
\usepackage{amsmath}
\usepackage{float}
\usepackage{bbm}

\usepackage{amsfonts}
\usepackage{bbm}
\usepackage{color}
\usepackage{latexsym}
\usepackage{caption}
\usepackage{subcaption}
\usepackage{times,txfonts}
\usepackage{color,soul}
\usepackage{listings}
\usepackage{bbold}
\usepackage{mathrsfs}
\usepackage[pdftex,plainpages=false,colorlinks=true,citecolor=blue,linkcolor=blue,urlcolor=blue,filecolor=green,bookmarksopen=true]{hyperref}
\usepackage{hyperref}

\begin{document}
	
	\title{Scalaron excitation by topological vortices in quadratic $f(R)$ gravity on a BTZ black hole background}
	
	\author{C. A. S. Almeida}
	\email{carlos@fisica.ufc.br}
	\affiliation{Universidade Federal do Cear\'a, Departamento de F\'{i}sica, 60455-760, Fortaleza, CE, Brazil}
	
	\author{F. C. E. Lima}
	\email{cleiton.estevao@fisica.ufc.br}
	\affiliation{Universidade Federal do Cear\'a, Departamento de F\'{i}sica, 60455-760, Fortaleza, CE, Brazil}

	
	\begin{abstract}
		In three spacetime dimensions, pure Einstein gravity admits no local propagating
degrees of freedom, yet nontrivial gravitational backgrounds such as the BTZ
black hole provide a natural arena to probe dynamical extensions of the theory.
In quadratic $f(R)$ gravity the Ricci scalar becomes a propagating degree of
freedom - the scalaron. We investigate how localized Maxwell-Higgs vortices excite this scalar mode in a static BTZ black-hole background. Working in the perturbative regime $\alpha \ll \ell^2$, the trace equation reduces to a massive Klein-Gordon equation for the curvature scalar sourced by the trace of the vortex
energy-momentum tensor. Using the Sturm-Liouville structure of the radial operator, we construct the corresponding Green function and obtain the curvature profile generated by an arbitrary localized source. The induced excitation exhibits a universal asymptotic decay $R(r) \sim r^{-(1+\nu)}$, independent of the detailed vortex structure. The scalar excitation is linearly stable, carries finite energy,
and produces parametrically suppressed backreaction, ensuring the smooth
recovery of the Einstein limit. These results provide a concrete realization of how higher-curvature corrections activate the unique local gravitational degree of freedom in three dimensions and how localized sources excite this scalar mode
in black-hole spacetimes.

			\end{abstract}

	\maketitle
	
\section{Introduction}

Gravity in three spacetime dimensions provides a remarkably clean arena for exploring fundamental aspects of gravitational dynamics. 
In pure Einstein gravity with negative cosmological constant, the Riemann tensor is algebraically determined by the Ricci tensor, and in vacuum the theory admits no local propagating degrees of freedom \cite{Deser1984,Carlip1998}. 
Nevertheless, nontrivial global solutions exist, most notably the Ba\~nados--Teitelboim--Zanelli (BTZ) black hole \cite{BTZ1992}, which has become a central laboratory for investigating black hole thermodynamics, quantum aspects of gravity, and holography in lower dimensions.

The absence of local graviton modes in three-dimensional Einstein gravity is not a generic property of all gravitational theories in $2+1$ dimensions. 
Higher-curvature extensions modify the dynamical content of the theory and may introduce genuine propagating degrees of freedom. 
One of the simplest and most extensively studied modifications is quadratic $f(R)$ gravity,
\begin{equation}
f(R) = R + \alpha R^2 ,
\end{equation}
which preserves general covariance while promoting curvature scalars to dynamical variables.
Through the scalar--tensor correspondence \cite{Whitt1984,Maeda1989,SotiriouFaraoni2010,DeFeliceTsujikawa2010}, 
quadratic $f(R)$ gravity is dynamically equivalent to Einstein gravity coupled to a massive scalar field. 
In $2+1$ dimensions, this scalar mode --- often referred to as the \emph{scalaron} \cite{Starobinsky1980} --- constitutes the unique local propagating gravitational degree of freedom.

In contrast to Einstein gravity, where the Ricci scalar is algebraically constrained by the trace of the energy--momentum tensor, quadratic $f(R)$ gravity renders the curvature scalar dynamical. 
The trace equation becomes a second-order differential equation of Klein--Gordon type, implying that localized sources with nonvanishing trace excite propagating curvature perturbations. 
The propagation of massive scalar fields in asymptotically AdS$_3$ spacetimes has been extensively studied \cite{Avis1978,LifschytzOrtiz1994,Birmingham2001}, 
and stability is governed by the Breitenlohner--Freedman bound \cite{BreitenlohnerFreedman1982}. 
These results provide a well-established framework within which the scalaron dynamics in the BTZ geometry can be consistently analyzed.

Topological defects offer natural localized sources for curvature excitations. 
The Maxwell--Higgs model supports vortex solutions characterized by finite energy and nontrivial winding number \cite{NielsenOlesen1973}. 
Such vortices possess a localized core region in which the trace of the energy--momentum tensor is generically nonzero. 
While vortex configurations in curved spacetimes have been investigated in various contexts, their role as sources of dynamical curvature modes in higher-curvature gravity has not been systematically examined in $2+1$ dimensions.

Although vortices in modified gravity have been studied previously, here we focus on a different aspect: the excitation of the scalar degree of freedom of quadratic $f(R)$ gravity. Therefore the purpose of the present work is to investigate how a localized Maxwell--Higgs vortex excites the scalar degree of freedom in quadratic $f(R)$ gravity in a BTZ black hole background. 
We focus on the perturbative regime $\alpha \ll \ell^2$, where $\ell$ denotes the AdS radius, and treat the higher-curvature corrections as small deviations from Einstein gravity. 
In this regime, the trace equation reduces to a massive Klein--Gordon equation for the Ricci scalar,
\begin{equation}
(\Box - m^2) R = - \frac{2\pi}{\alpha} T,
\end{equation}
with effective mass $m^2 = \frac{1}{4\alpha}$. The scalaron therefore propagates as a massive field in the fixed BTZ geometry, sourced by the trace of the vortex energy--momentum tensor.

The small-$\alpha$ regime admits a natural interpretation as the leading order of a controlled effective field theory expansion of gravity \cite{Donoghue1994}. 
The perturbative treatment is governed by a dimensionless parameter proportional to $\alpha R$, ensuring that nonlinear corrections remain suppressed for sufficiently localized sources.

Exploiting the Sturm--Liouville structure of the radial scalar operator in the BTZ background, we construct explicitly the associated Green function subject to physically appropriate boundary conditions: regularity at the horizon and normalizability at spatial infinity. 
This construction parallels the analysis of scalar field propagation in the BTZ geometry \cite{LifschytzOrtiz1994,Birmingham2001}. 
We show that, outside the vortex core, the induced curvature perturbation exhibits a universal asymptotic behavior
\begin{equation}
R(r) \sim r^{-(1+\nu)}, 
\qquad 
\nu = \sqrt{1 + m^2 \ell^2},
\end{equation}
independent of the detailed structure of the vortex core. 
The exponent $\nu$ coincides with the standard AdS mass--dimension relation for scalar fields \cite{Avis1978,Witten1998}, reflecting the underlying asymptotically AdS$_3$ geometry.

We further demonstrate that the scalar excitation is linearly stable, since $m^2 > 0$ lies well above the Breitenlohner--Freedman bound \cite{BreitenlohnerFreedman1982}, and carries finite total energy. 
Using the quadratic scalar--tensor form of the effective action, the energy can be consistently defined within the framework of higher-curvature gravity \cite{DeserTekin2003}. 
In the perturbative regime $\alpha \ll \ell^2$, the scalar energy is parametrically suppressed relative to the BTZ mass parameter, ensuring that backreaction effects remain small and that the Einstein limit is recovered smoothly as $\alpha \to 0$.

The mechanism identified here provides a transparent realization of how higher-curvature corrections activate local gravitational dynamics in three dimensions. 
Although our analysis is restricted to quadratic $f(R)$ gravity, the qualitative picture is expected to persist in broader classes of higher-derivative theories in which curvature scalars become dynamical. 
The $2+1$-dimensional setting offers a particularly controlled framework in which the unique propagating gravitational degree of freedom can be isolated and analyzed analytically.

This paper is organized as follows. 
In Sec.~II we review quadratic $f(R)$ gravity and derive the linearized trace equation. 
In Sec.~III we analyze the scalar operator in the BTZ background and construct the radial Green function. 
Section IV is devoted to the curvature profile generated by a localized vortex and its universal asymptotic behavior. 
In Sec.~V we analyze stability and energetic properties of the scalar excitation. 
Section VI discusses the consistency of the perturbative regime and the smooth recovery of the Einstein limit. 
We conclude in Sec.~VII.


\section{Quadratic $f(R)$ gravity and the scalar degree of freedom}

Modified gravity theories of the $f(R)$ type constitute one of the simplest extensions of Einstein gravity and have been extensively studied in cosmological and astrophysical contexts (see, e.g., \cite{SotiriouFaraoni2010, DeFeliceTsujikawa2010}). In particular, quadratic gravity of the form
\begin{equation}
f(R)=R+\alpha R^2 ,
\label{fR_model}
\end{equation}
introduces higher-curvature corrections while preserving general covariance.

In three spacetime dimensions, pure Einstein gravity exhibits no local propagating degrees of freedom \cite{DeserJackiw1984, Carlip1998}. This follows from the fact that the Riemann tensor is algebraically determined by the Ricci tensor, and in vacuum the theory is locally trivial. However, the addition of higher-curvature terms alters the dynamical content of the theory. In particular, quadratic $f(R)$ gravity is dynamically equivalent to Einstein gravity coupled to a scalar field via a conformal transformation \cite{SotiriouFaraoni2010}. In $2+1$ dimensions, this scalar mode represents the only propagating gravitational degree of freedom.

We consider the action
\begin{equation}
S=\int d^3x \sqrt{-g}
\left[
\frac{1}{16\pi}\left(R+\alpha R^2\right)
+\mathcal L_{MH}
\right],
\label{action_total}
\end{equation}
where $\mathcal L_{MH}$ denotes the Maxwell--Higgs Lagrangian supporting vortex solutions \cite{NielsenOlesen1973}. Variation with respect to the metric yields the field equations
\begin{equation}
(1+2\alpha R)R_{\mu\nu}
-\frac{1}{2}g_{\mu\nu}(R+\alpha R^2)
+2\alpha\left(g_{\mu\nu}\Box-\nabla_\mu\nabla_\nu\right)R
=8\pi T_{\mu\nu}.
\label{field_eq_full}
\end{equation}

The field equations (\ref{field_eq_full}) follow from the standard metric variation of the $f(R)$ action 
(see, e.g., Refs.~\cite{Whitt1984,Maeda1989,SotiriouFaraoni2010,DeFeliceTsujikawa2010}). 
They reduce to Einstein equations in the limit $\alpha \to 0$.

Taking the trace of Eq.~(\ref{field_eq_full}) leads to
\begin{equation}
R+2\alpha R^2-4\alpha \Box R
=8\pi T,
\label{trace_full}
\end{equation}
where $T=g^{\mu\nu}T_{\mu\nu}$.

Equation (\ref{trace_full}) makes explicit that, unlike in Einstein gravity where R is algebraically
determined by T , the curvature scalar becomes dynamical in quadratic f (R) gravity.
This feature is well known in four dimensions \cite{SotiriouFaraoni2010,DeFeliceTsujikawa2010},  but acquires special significance
in 2 + 1 dimensions, where it corresponds to the emergence of the only propagating
gravitational mode.


\subsection*{Linearized regime}

We now focus on the perturbative regime
\begin{equation}
\alpha \ll \ell^2,
\end{equation}
where $\ell$ denotes the AdS radius of the background geometry introduced below. In this regime we assume
\begin{equation}
\alpha R \ll 1,
\end{equation}
so that quadratic terms in $R$ can be neglected in Eq.~(\ref{trace_full}). The trace equation then reduces to
\begin{equation}
\Box R - m^2 R
= -\frac{2\pi}{\alpha} T,
\label{trace_linear}
\end{equation}
with effective mass
\begin{equation}
m^2=\frac{1}{4\alpha}.
\label{mass_scalar}
\end{equation}

Equation (\ref{trace_linear}) makes explicit that, at linear order in $\alpha$, the curvature scalar 
propagates as a massive scalar field in the fixed background geometry. 
This reduction to a Klein-Gordon-type equation is a standard feature of quadratic $f(R)$ gravity 
in its scalar-tensor representation \cite{Whitt1984,SotiriouFaraoni2010}. 

On the other hand, Equation~(\ref{trace_linear}) has the form of a Klein-Gordon equation for a massive scalar field propagating in a curved background and sourced by the trace of the energy-momentum tensor. The positivity of $m^2$ for $\alpha>0$ ensures the absence of tachyonic instabilities in the flat-space limit.

It is convenient to introduce the scalar field
\begin{equation}
\Phi \equiv f_R = 1+2\alpha R,
\end{equation}
which corresponds to the scalar degree of freedom in the scalar-tensor representation of $f(R)$ gravity \cite{SotiriouFaraoni2010}. In the linear regime, $\Phi-1 \propto R$, and the quadratic part of the action reduces to
\begin{equation}
S_{\rm eff}
\sim
\int d^3x \sqrt{-g}
\left[
\frac{1}{2}(\nabla R)^2
-\frac{1}{2}m^2 R^2
\right],
\label{scalar_action}
\end{equation}
confirming that the curvature perturbation behaves as a massive scalar excitation.

This quadratic effective action captures the propagating scalaron mode and 
provides the appropriate starting point for analyzing stability and energetic properties. 
The interpretation of the $R^2$ correction as introducing a genuine scalar degree of freedom 
goes back to the seminal work of Starobinsky \cite{Starobinsky1980}.

In the following sections we investigate how localized Maxwell-Higgs vortices, whose energy-momentum tensor generically has a non-vanishing trace in the core region, act as sources for this scalar degree of freedom in a BTZ black hole background.


\section{BTZ background and radial scalar operator}

In three spacetime dimensions with negative cosmological constant, the most general static and circularly symmetric vacuum solution of Einstein gravity is the Banados-Teitelboim--Zanelli (BTZ) black hole \cite{BTZ1992}. Its metric can be written as
\begin{equation}
ds^2 = -h_0(r) dt^2 + \frac{dr^2}{h_0(r)} + r^2 d\phi^2,
\label{BTZ_metric}
\end{equation}
with
\begin{equation}
h_0(r) = \frac{r^2}{\ell^2} - M ,
\label{BTZ_h}
\end{equation}
where $\ell$ denotes the AdS radius and $M$ is the dimensionless mass parameter. The horizon radius is determined by
\begin{equation}
h_0(r_h)=0 
\quad \Rightarrow \quad 
r_h=\ell \sqrt{M}.
\label{horizon_radius}
\end{equation}

The existence of static Maxwell-Higgs vortex configurations in 
three-dimensional curved spacetimes has been established in the 
literature for both Minkowski and AdS$_{3}$ backgrounds. Edery 
\cite{Edery2021} demonstrated that non-singular vortices with 
positive mass can be consistently supported in $(2+1)$-dimensional 
Einstein gravity with an AdS$_{3}$ background, providing the 
geometric foundation for the present analysis. In the context of 
black hole spacetimes, Lima, Moreira and Almeida \cite{Lima2023} 
constructed explicit numerical solutions for the vortex field 
variables $g(r)$ and $a(r)$ in the background of a 
three-dimensional black hole in Einstein gravity, confirming that 
static vortex configurations are stable outside the event horizon 
and that the field equations admit well-behaved solutions 
satisfying the topological boundary conditions $g(0)=0$, 
$g(\infty)=\nu$, $a(0)=0$, and $a(\infty)=-\beta$. In the 
perturbative regime $\alpha \ll \ell^{2}$ considered throughout 
this work, the higher-curvature corrections to the background 
geometry are parametrically small, and the vortex existence 
problem reduces at leading order to the Einstein gravity case 
analyzed in those references. The present paper therefore takes 
the existence of the vortex configuration as established and 
focuses on the novel phenomenon of scalaron excitation induced 
by its energy--momentum tensor.

In the perturbative regime considered here, we treat the BTZ geometry as a fixed background and focus on the excitation of the scalar curvature perturbation governed by the linearized trace equation (\ref{trace_linear}). Since the vortex configuration is static and circularly symmetric, we assume that the curvature perturbation depends only on the radial coordinate,
\begin{equation}
R = R(r).
\end{equation}

\subsection*{Radial form of the scalar equation}

For the metric (\ref{BTZ_metric}), the d'Alembertian acting on a radial scalar reduces to
\begin{equation}
\Box R
=
\frac{1}{\sqrt{-g}}
\partial_r \!\left(
\sqrt{-g} \, g^{rr} \partial_r R
\right).
\end{equation}
Since $\sqrt{-g}=r$ and $g^{rr}=h_0(r)$, this becomes
\begin{equation}
\Box R
=
\frac{1}{r}
\frac{d}{dr}
\left[
r h_0(r) \frac{dR}{dr}
\right].
\label{box_radial}
\end{equation}

The structure of the radial Klein-Gordon operator in the BTZ background has been 
extensively investigated in the literature in the context of scalar field propagation 
and quasinormal modes (see, e.g., Refs.~\cite{LifschytzOrtiz1994,Birmingham2001}). 
The present problem differs only by the interpretation of the scalar as a curvature excitation.

Substituting (\ref{box_radial}) into the linearized trace equation (\ref{trace_linear}), we obtain the radial equation
\begin{equation}
\frac{1}{r}
\frac{d}{dr}
\left[
r h_0(r) \frac{dR}{dr}
\right]
- m^2 R
=
- \frac{2\pi}{\alpha} T(r).
\label{radial_eq_compact}
\end{equation}

Equation (\ref{radial_eq_compact}) can be expanded explicitly as
\begin{equation}
\left(
\frac{r^2}{\ell^2}-M
\right) R''(r)
+
\left(
\frac{3r}{\ell^2}-\frac{M}{r}
\right) R'(r)
-
m^2 R(r)
=
- \frac{2\pi}{\alpha} T(r).
\label{radial_eq_expanded}
\end{equation}

\subsection*{Self-adjoint structure}

Equation (\ref{radial_eq_compact}) reveals that the radial operator can be written in Sturm-Liouville form,
\begin{equation}
\mathcal{L}R
=
\frac{1}{r}
\frac{d}{dr}
\left[
r h_0(r) R'
\right]
-
m^2 R,
\label{SL_operator}
\end{equation}
with weight function $w(r)=r$. This self-adjoint structure plays a central role in the construction of the Green function in the following section. In particular, it ensures the existence of a complete set of homogeneous solutions and guarantees that the scalar excitation can be represented as a convolution of the source $T(r)$ with the appropriate radial Green kernel.

\subsection*{Homogeneous equation and asymptotic behavior}

Outside the vortex core, where $T(r)=0$, the scalar curvature satisfies the homogeneous equation
\begin{equation}
\frac{1}{r}
\frac{d}{dr}
\left[
r h_0(r) R'
\right]
-
m^2 R
=
0.
\label{homogeneous_eq}
\end{equation}

Two physically relevant independent solutions can be characterized by their boundary behavior:

\begin{itemize}
\item[(i)] A solution regular at the horizon $r=r_h$;
\item[(ii)] A solution decaying at spatial infinity.
\end{itemize}

In the asymptotic region $r\gg r_h$, the metric function behaves as $h_0(r)\sim r^2/\ell^2$, and Eq.~(\ref{homogeneous_eq}) reduces to
\begin{equation}
R'' + \frac{3}{r} R' - \frac{m^2 \ell^2}{r^2} R \simeq 0.
\label{asymptotic_eq}
\end{equation}

Seeking power-law solutions $R \sim r^{-p}$ leads to
\begin{equation}
p(p-2)=m^2 \ell^2.
\end{equation}
Defining
\begin{equation}
\nu = \sqrt{1 + m^2 \ell^2},
\label{nu_def}
\end{equation}
the asymptotic behavior becomes
\begin{equation}
R(r) \sim r^{-(1+\nu)}.
\label{asymptotic_solution}
\end{equation}

For $\alpha>0$, one has $m^2>0$ and therefore $\nu>1$, implying a rapid decay of the scalar excitation at large distances. This behavior will be central to the analysis of stability and energy finiteness in subsequent sections.

The exponent $\nu$ coincides with the standard AdS mass--dimension relation 
for scalar fields in asymptotically AdS$_3$ spacetimes \cite{Avis1978,Witten1998}. 
This reflects the fact that, at linear order, the scalaron behaves as an ordinary 
massive scalar field propagating on the BTZ geometry.

\section{Green function and scalar excitation by the vortex}

For definiteness, one may model the trace of the vortex energy--momentum tensor 
as a smooth localized function with compact support around $r \sim r_v$, for example
\[
T(r) \sim T_0 \exp\left[-(r-r_v)^2/\sigma^2\right],
\]
with $\alpha \ll r_v$.
The subsequent analysis, however, depends only on the integrated effective charge
$Q_{\rm eff}$ and not on the detailed shape of $T(r)$.

In this section we construct the radial Green function associated with the operator
\begin{equation}
\mathcal{L}R
=
\frac{1}{r}
\frac{d}{dr}
\left[
r h_0(r) R'
\right]
-
m^2 R ,
\label{L_operator_repeat}
\end{equation}
and determine the curvature perturbation generated by a localized vortex source.

\subsection*{Construction of the radial Green function}

The Green function $G(r,r')$ is defined by
\begin{equation}
\mathcal{L} G(r,r')
=
\frac{\delta(r-r')}{r},
\label{Green_def}
\end{equation}
where the factor $1/r$ reflects the Sturm-Liouville weight of the operator.

Let $R_1(r)$ and $R_2(r)$ be two independent solutions of the homogeneous equation
\begin{equation}
\mathcal{L}R=0,
\label{homogeneous_repeat}
\end{equation}
with the following physical boundary conditions:

\begin{itemize}
\item $R_1(r)$ is regular at the horizon $r=r_h$;
\item $R_2(r)$ decays at spatial infinity.
\end{itemize}

Given the self-adjoint structure of $\mathcal{L}$, the Green function can be written as
\begin{equation}
G(r,r')
=
\frac{1}{W(r')}
\begin{cases}
R_1(r)\,R_2(r') , & r<r', \\[4pt]
R_1(r')\,R_2(r) , & r>r',
\end{cases}
\label{Green_general}
\end{equation}
where the Wronskian is defined as
\begin{equation}
W(r)
=
r h_0(r)
\left(
R_1 R_2' - R_2 R_1'
\right),
\label{Wronskian}
\end{equation}
and is independent of $r$ up to normalization.

The representation (\ref{Green_general}) follows from the general Green function construction 
for second-order self-adjoint Sturm-Liouville operators. 
Its validity relies on the regularity of the homogeneous solutions and on the 
constancy of the Wronskian up to normalization.

\subsection*{Scalar profile generated by a localized vortex}

The curvature perturbation generated by a vortex with trace source $T(r)$ is given by
\begin{equation}
R(r)
=
- \frac{2\pi}{\alpha}
\int_{r_h}^{\infty}
G(r,r')\,T(r')\, r' dr'.
\label{R_integral}
\end{equation}

We assume that the vortex is localized outside the horizon, with $r_v > r_h$ and finite core size of order $r_v$. 
No extreme hierarchy between $r_v$ and $r_h$ is required for the analysis.

In the asymptotic region $r\gg r_h$, the homogeneous solution decaying at infinity behaves as
\begin{equation}
R_2(r) \sim r^{-(1+\nu)},
\qquad
\nu=\sqrt{1+m^2\ell^2}.
\label{R2_asymptotic}
\end{equation}

Since $R_1(r')$ is finite at the horizon and smooth in the region $r'\sim r_v$, it can be approximated by a constant within the vortex core. Assuming that the homogeneous solution regular at the horizon varies slowly over the vortex core scale, Eq.~(\ref{R_integral}) reduces to
\begin{equation}
R(r)
\sim
\frac{Q_{\rm eff}}{4\alpha (1+\nu)}\, r^{-(1+\nu)},
\label{R_asymptotic}
\end{equation}
where we have introduced the effective scalar charge
\begin{equation}
Q_{\rm eff}
=
8\pi
\int_{r_h}^{\infty}
\frac{r'^{2+\nu}}{h_0(r')}
T(r')\, dr'.
\label{Qeff_def}
\end{equation}

Equation (\ref{R_asymptotic}) shows that the detailed structure of the vortex core affects only 
the overall normalization through $Q_{\rm eff}$, while the radial decay is universal and 
entirely determined by the effective mass $m^2=1/(4\alpha)$ and the AdS curvature scale $\ell$. 
In this sense, the scalar excitation is insensitive to microscopic vortex details 
and reflects only the infrared properties of the background geometry.


The effective scalar charge defined in Eq.~(35), contains an explicit factor of $1/h_0(r')$, which vanishes at the horizon $r'=r_h$. It is therefore important to verify that the integral is well defined.

This issue is resolved by the physical localization of the vortex source. The Maxwell--Higgs vortex is assumed to be centered at a radius $r_v$ well outside the black hole horizon, with a finite core size of order $r_v$. Consequently, the trace of the energy--momentum tensor $T(r')$ has compact (or exponentially suppressed) support around $r' \sim r_v$ and vanishes identically in a neighborhood of the horizon. As a result, the integration region that contributes to $Q_{\mathrm{eff}}$ never probes the near-horizon behavior of $h_0(r')$, and the apparent singularity at $r'=r_h$ is physically irrelevant.

More explicitly, since $T(r')=0$ for $r' \lesssim r_v-\Delta r$ with $\Delta r \ll r_v$, the lower limit of the effective integration domain can be taken as $r' \gtrsim r_v$, where $h_0(r')$ is strictly positive and smooth. In this region, the integrand is regular and the integral converges trivially. Therefore, the effective scalar charge $Q_{\mathrm{eff}}$ is finite and well defined for vortices localized outside the horizon.

This assumption is physically natural, since a static vortex configuration cannot be supported arbitrarily close to the horizon without being absorbed by the black hole. 

Outside the vortex core, the curvature profile decays as $R(r) \sim r^{-(1+\nu)}$. This behavior is illustrated in Fig.~1 for representative values of the quadratic coupling $\alpha$, confirming the universal nature of the asymptotic decay.
\begin{figure}
  \centering
  \includegraphics[width=0.7\linewidth]{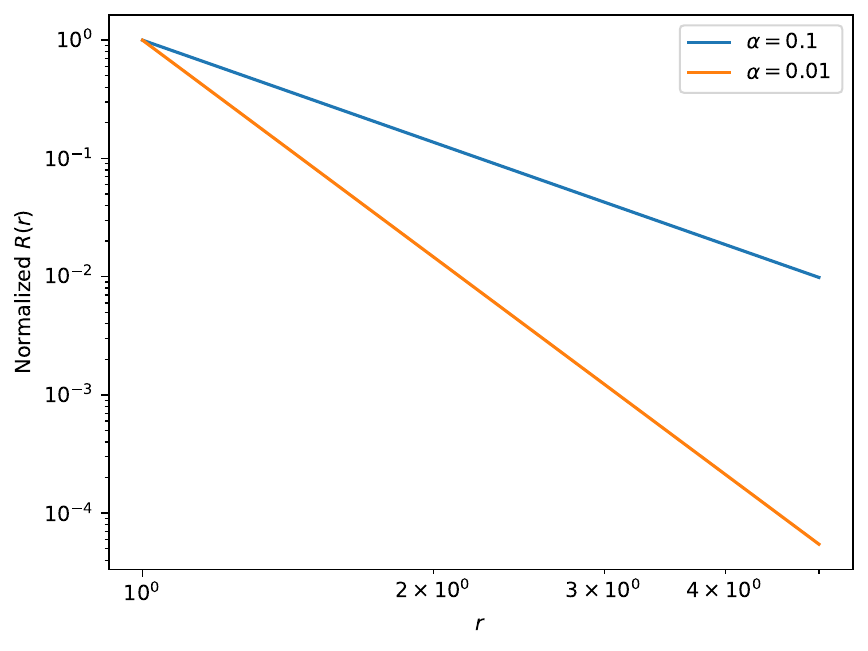}
  \caption{Log--log plot of the normalized scalar curvature profile $R(r)$ outside the
vortex core for representative values of the quadratic coupling parameter $\alpha$.
The curves correspond to $\alpha = 0.1$ and $\alpha = 0.01$, with the AdS radius fixed
to $\ell = 1$ and the horizon radius set to $r_h = 1$. The curvature profile has been
normalized by its value at the vortex core radius, $R(r_v)$, in order to highlight the
universal radial decay. The straight-line behavior in the log--log plot confirms the
asymptotic power-law falloff $R(r) \sim r^{-(1+\nu)}$, where
$\nu = \sqrt{1 + \ell^2/(4\alpha)}$. Smaller values of $\alpha$ correspond to larger
effective scalar mass and thus to stronger localization of the scalar excitation.
}
\end{figure}

\subsection*{Universal features of the scalar excitation}

The scalar curvature perturbation induced by the vortex therefore exhibits the following properties:

\begin{itemize}
\item[(i)] It is localized, decaying as a power law controlled by $\nu$;
\item[(ii)] It vanishes smoothly in the Einstein limit $\alpha \to 0$ (for which $\nu \to \infty$);
\item[(iii)] Its existence is mandatory whenever the trace $T(r)$ is non-zero.
\end{itemize}

Thus, in quadratic $f(R)$ gravity in $2+1$ dimensions, topological vortices act as localized sources for the only local gravitational degree of freedom. The excitation is entirely determined by the Green kernel of the BTZ background and is independent of the microscopic details of the vortex, apart from the overall scalar charge $Q_{\rm eff}$.


\section{Stability and energetic properties of the scalar excitation}

In this section we analyze the stability and energy content of the scalar curvature excitation induced by the vortex. These properties are essential for assessing the physical consistency of the perturbative regime considered throughout this work.

\subsection*{Linear stability}

The curvature perturbation satisfies the Klein-Gordon-type equation
\begin{equation}
\Box R - m^2 R = - \frac{2\pi}{\alpha} T,
\label{KG_equation_repeat}
\end{equation}
with effective mass
\begin{equation}
m^2=\frac{1}{4\alpha}.
\end{equation}

For $\alpha>0$, one has $m^2>0$, and therefore the scalar mode is non-tachyonic in the flat-space limit. Since the BTZ geometry is locally AdS$_3$, stability in the asymptotic region is governed by the Breitenlohner-Freedman (BF) criterion \cite{BreitenlohnerFreedman1982}. In three dimensions, the corresponding bound reads
\begin{equation}
m^2 \ge -\frac{1}{\ell^2}.
\end{equation}
Because $m^2=1/(4\alpha)>0$, the scalar mass lies well above this threshold. The scalar excitation is therefore linearly stable in the asymptotically AdS$_3$ background.


\subsection*{Energy of the scalar excitation}

In the scalar-tensor representation of quadratic $f(R)$ gravity \cite{SotiriouFaraoni2010}, the curvature perturbation behaves as a massive scalar field. In the linear regime, the quadratic part of the effective action takes the form
\begin{equation}
S_{\rm eff}
\sim
\int d^3x \sqrt{-g}
\left[
\frac{1}{2}(\nabla R)^2
-\frac{1}{2} m^2 R^2
\right],
\label{effective_action_repeat}
\end{equation}
up to an overall normalization factor, which does not affect the convergence analysis.

For static and circularly symmetric configurations, the corresponding energy density is
\begin{equation}
\rho_R
=
\frac{1}{2} h_0(r) \left( R'(r) \right)^2
+
\frac{1}{2} m^2 R(r)^2.
\label{energy_density}
\end{equation}

The total energy of the scalar excitation is therefore
\begin{equation}
E_R
=
2\pi
\int_{r_h}^{\infty}
\left[
\frac{1}{2} h_0(r) (R')^2
+
\frac{1}{2} m^2 R^2
\right]
r\, dr.
\label{total_energy}
\end{equation}

The definition of conserved energy in higher-curvature gravity theories has been 
discussed in detail in Ref.~\cite{DeserTekin2003}. 
In the present perturbative regime, the scalar sector reduces consistently to 
the quadratic scalar-tensor form (\ref{effective_action_repeat}), and the energy functional (\ref{total_energy}) 
provides a reliable measure of the excitation amplitude.

Using the asymptotic behavior derived in Sec.~IV,
\begin{equation}
R(r) \sim r^{-(1+\nu)},
\qquad
\nu=\sqrt{1+m^2\ell^2},
\end{equation}
one finds
\begin{equation}
R'(r) \sim r^{-(2+\nu)}.
\end{equation}

Because $\nu>0$ for $\alpha>0$, the integral (\ref{total_energy}) converges strongly at infinity. Near the horizon, the regularity of $R(r)$ ensures finiteness of the energy density. Hence,
\begin{equation}
E_R < \infty.
\end{equation}

\subsection*{Physical interpretation}

The finiteness of $E_R$ establishes that the scalar excitation is energetically localized 
and does not generate infrared divergences in the asymptotically AdS$_3$ geometry. 
Moreover, its parametric suppression in the regime $\alpha \ll \ell^2$ ensures that 
the scalar energy remains much smaller than the BTZ mass parameter $M$, 
validating the treatment of the background geometry as fixed at leading order.

The excitation is therefore dynamically consistent and energetically well-behaved. In quadratic $f(R)$ gravity in $2+1$ dimensions, topological vortices induce a localized, finite-energy perturbation of the propagating curvature mode, without triggering instabilities or large backreaction effects.


\section{Consistency of the perturbative regime and Einstein limit}

In this section we examine the self-consistency of the perturbative treatment adopted throughout the paper. Two aspects must be addressed: (i) the validity of the linearization of the trace equation, and (ii) the smallness of the scalar backreaction relative to the background geometry.

\subsection*{Validity of the linear approximation}

The full trace equation,
\begin{equation}
R + 2\alpha R^{2} - 4\alpha \Box R = 8\pi T ,
\end{equation}
was linearized under the assumption $\alpha R \ll 1$. Although this condition is verified \emph{a posteriori}, its validity is controlled by a well-defined dimensionless parameter in the perturbative regime. Indeed, since the linearized equation for $R$ is elliptic and contains a positive mass term, the curvature profile does not exhibit amplification outside the source region. As a consequence, the maximum of $R(r)$ occurs within the vortex core, at $r \sim r_v$.

Using the asymptotic solution derived in Sec.~IV, the maximal amplitude of the curvature perturbation can be estimated as
\begin{equation}
R_{\mathrm{max}} \sim \frac{T_0\, r_v^{2}}{4\alpha(1+\nu)} ,
\end{equation}
where $T_0$ characterizes the typical magnitude of the trace of the energy--momentum tensor inside the vortex core and
\begin{equation}
\nu = \sqrt{1+\frac{\ell^{2}}{4\alpha}} .
\end{equation}
In the perturbative regime $\alpha \ll \ell^{2}$, one has $\nu \gg 1$, and the linearization condition takes the form
\begin{equation}
\alpha R_{\mathrm{max}} \sim \frac{T_0 r_v^{2}}{\nu}
\sim \frac{T_0 r_v^{2} \sqrt{\alpha}}{\ell} \ll 1 .
\end{equation}
This identifies a small dimensionless parameter that controls the perturbative expansion and ensures that the nonlinear terms of order $\mathcal{O}(\alpha R^{2})$ are systematically suppressed with respect to the linear contribution.

Therefore, for localized sources and sufficiently small values of $\alpha$, the linearized trace equation provides a self-consistent description throughout spacetime. The approximation adopted in this work is thus not merely heuristic, but corresponds to the leading order of a controlled expansion in the parameter $\alpha R$.

This scaling behavior is consistent with the effective field theory interpretation 
of higher-curvature gravity, where $\alpha$ controls the strength of higher-derivative 
corrections and the expansion is organized in powers of $\alpha R$ 
(see Ref.~\cite{Donoghue1994}). 
The perturbative treatment adopted here therefore corresponds to the leading order 
of a controlled expansion.

\subsection*{Smallness of the scalar backreaction}

In order to assess the consistency of treating the BTZ geometry as a fixed background,
it is necessary to verify that the scalar excitation induced by the vortex carries a
negligible amount of energy compared to the mass parameter $M$ of the black hole.
Since the curvature perturbation behaves as a massive scalar degree of freedom in the
linearized regime, its energy can be computed from the quadratic part of the effective
action.

For static and circularly symmetric configurations, the energy associated with the
scalar mode is given by
\begin{equation}
E_R = 2\pi \int_{r_h}^{\infty}
\left[
\frac{1}{2} h_0(r)\, (R'(r))^2
+ \frac{1}{2} m^2 R(r)^2
\right] r\, dr .
\end{equation}
As shown in the previous subsection, the curvature profile outside the vortex core
decays as
\begin{equation}
R(r) \sim \mathcal{A}\, r^{-(1+\nu)},
\qquad
\nu = \sqrt{1+m^2\ell^2} ,
\end{equation}
with an amplitude $\mathcal{A}$ determined by the effective scalar charge
$Q_{\mathrm{eff}}$. Substituting this asymptotic behavior into the energy functional,
both the gradient and mass terms contribute integrands that scale at large radius as
$r^{-2\nu-1}$. Consequently, the energy integral converges strongly at infinity for
$\nu>0$, confirming that the scalar excitation carries finite total energy.

To estimate the magnitude of this energy, it is sufficient to perform a parametric
evaluation of the integral. Since the integrand is strongly suppressed at large radius,
the dominant contribution arises from the lower end of the asymptotic region, namely
from radii of order the vortex core scale $r \sim r_v$. Evaluating the integrand at $r \sim r_v$, where the curvature amplitude is maximal, yields the parametric estimate
\begin{equation}
E_R \sim \mathcal{A}^2\, r_v^{-2\nu} ,
\end{equation}
up to numerical factors of order unity.

The amplitude $\mathcal{A}$ is itself suppressed by the large mass parameter $\nu$ and
by the localization of the source. Using the relation between $\mathcal{A}$ and
$Q_{\mathrm{eff}}$, one finds that the scalar energy can be expressed parametrically as
\begin{equation}
E_R \sim \mathcal{C}\,
\left( \frac{r_h}{r_v} \right)^{2\nu} ,
\end{equation}
where $\mathcal{C}$ collects powers of $\alpha$, $T_0$, and $r_v$, as well as numerical
coefficients that are not relevant for the present discussion. This expression should be
understood as a parametric estimate rather than an exact closed-form result; its purpose
is to exhibit the strong suppression mechanism governing the scalar backreaction. This scaling follows from the fact that the amplitude A is proportional to the value of the regular homogeneous solution evaluated near the horizon, which introduces a factor scaling as $r_h^{\nu}$.

In the perturbative regime $ \alpha \ll \ell^2$, one has $\nu \gg 1$. 
Since the vortex is localized outside the horizon, $r_v > r_h$, the ratio $r_h/r_v$ is strictly smaller than unity. 
It is therefore convenient to rewrite the suppression factor as 
\begin{equation*}
\left(\frac{r_h}{r_v}\right)^{2\nu}
=
\exp\!\left[ 2\nu \ln\!\left(\frac{r_h}{r_v}\right) \right].
\end{equation*}
where $\ln(r_h/r_v) < 0$. 

For large $\nu$ this produces an exponential suppression of the scalar energy. 
As a result, the scalar contribution to the total energy remains parametrically much smaller than the BTZ mass parameter,
$E_R \ll M.$

We therefore conclude that the scalar excitation induced by the vortex produces a
negligible backreaction on the background geometry in the regime considered here. This
justifies treating the BTZ solution as a fixed background at leading order and confirms
the internal consistency of the perturbative approach.


\subsection*{Smooth Einstein limit}

Thus, the scalar excitation decouples smoothly and the theory reduces to three-dimensional Einstein gravity, in which no propagating degrees of freedom are present. The mechanism described in this work therefore does not introduce discontinuities or pathologies in the GR limit.

\medskip

In the limit $\alpha \to 0$, the scalar mass diverges and the scalaron decouples 
from the low-energy dynamics. 
The theory reduces smoothly to three-dimensional Einstein gravity, 
which contains no local propagating degrees of freedom. 
No discontinuity or pathology arises in this limit, and the perturbative 
construction remains internally consistent throughout the parameter range considered.


\section{Conclusions}

In this work we have investigated the behavior of curvature excitations induced by a
localized vortex in three-dimensional gravity with a quadratic curvature correction.
Working within the perturbative regime $\alpha R \ll 1$, we have shown that the trace
equation reduces to an effective massive scalar equation, whose solutions are fully
determined by the properties of the source and by the effective mass scale introduced
by the higher-derivative term.

We derived the scalar curvature profile generated by the vortex and showed that, outside
the core, it exhibits a universal power-law decay governed by a single parameter
$\nu = \sqrt{1 + \ell^{2}/(4\alpha)}$. This behavior is independent of the microscopic
details of the vortex and reflects the massive nature of the scalar mode associated with
the $R^{2}$ correction. The effective scalar charge sourcing this mode was shown to be
finite and well defined for vortices localized away from the black hole horizon.

We further analyzed the energetic content of the scalar excitation and demonstrated that
its total energy is finite and parametrically suppressed in the perturbative regime.
For $\alpha \ll \ell^{2}$, corresponding to $\nu \gg 1$, the scalar energy is strongly
suppressed and remains negligible compared to the BTZ mass parameter. This establishes
the smallness of the scalar backreaction and justifies treating the background geometry
as fixed at leading order.

Our results show that localized matter sources in higher-derivative gravity can induce
long-range curvature profiles without significantly affecting the background spacetime.
This provides a controlled setting in which the physical implications of higher-curvature
terms can be isolated and studied analytically. Possible extensions of this work include
the analysis of rotating solutions, time-dependent sources, and the generalization to
other classes of higher-curvature theories.

The 2+1 dimensional setting thus provides a minimal laboratory where higher-curvature gravity activates a single propagating geometric mode whose response to localized sources can be computed explicitly.



\section*{Acknowledgments}
\hspace{0.5cm}The authors are grateful to the Conselho Nacional de Desenvolvimento Cient\'{i}ico e Tecnol\'{o}gico (CNPq). F. C. E. Lima and C. A. S. Almeida are supported, respectively, for grants No. 171048/2023-7 (CNPq/PDJ) and 309553/2021-0 (CNPq/PQ). C. A. S. Almeida acknowledges Funda\c{c}\~{a}o Cearense de Apoio ao Desenvolvimento Cient\'{i}fico e Tecnol\'{o}gico (FUNCAP) for their valuable support through the Project UNI-00210-00230.01.00/23.

\section*{CONFLICTS OF INTEREST/COMPETING INTEREST}

The authors declared that there is no conflict of interest in this manuscript. 

\section*{DATA AVAILABILITY}

No data was used for the research described in this article.	

\appendix
\section*{Appendix A: Parametric estimates and large-$\nu$ regime}

In this appendix we provide the intermediate steps leading to the parametric
estimates used in Sec.~VI. The purpose is to clarify the scaling arguments,
rather than to compute exact numerical coefficients.

\subsection*{A.1 Asymptotic scalar profile}

Outside the vortex core, the trace equation reduces to
\begin{equation}
(\Box - m^2) R = -\frac{2\pi}{\alpha} T,
\end{equation}
with
\begin{equation}
m^2 = \frac{1}{4\alpha}.
\end{equation}

In the asymptotic region of the BTZ background, the radial solution behaves as
\begin{equation}
R(r) \sim A\, r^{-(1+\nu)},
\qquad
\nu = \sqrt{1 + m^2 \ell^2}.
\end{equation}

The constant $A$ is determined by matching to the source region.
Since the vortex is localized around $r \sim r_v$ with finite core size,
dimensional analysis implies the parametric form
\begin{equation}
R(r) \sim \frac{Q_{\rm eff}}{r^{1+\nu}},
\end{equation}
up to factors of order unity.

Evaluating at $r \sim r_v$ gives an estimate for the maximal curvature amplitude,
\begin{equation}
R_{\max} \sim \frac{Q_{\rm eff}}{r_v^{1+\nu}}.
\end{equation}

\subsection*{A.2 Large-$\nu$ behavior in the perturbative regime}

In the perturbative regime $\alpha \ll \ell^2$, one has
\begin{equation}
m^2 \ell^2 = \frac{\ell^2}{4\alpha} \gg 1.
\end{equation}

Therefore,
\begin{equation}
\nu = \sqrt{1 + \frac{\ell^2}{4\alpha}}
\simeq
\frac{\ell}{2\sqrt{\alpha}}
\gg 1.
\end{equation}

The scalaron is thus very massive, and the curvature profile decays rapidly outside the vortex core,
\begin{equation}
R(r) \sim r^{-(1+\nu)}.
\end{equation}

This rapid decay underlies the strong suppression effects discussed in the main text.

\subsection*{A.3 Parametric estimate of the scalar energy}

The scalar energy in the linear regime is
\begin{equation}
E_R = 2\pi \int_{r_h}^{\infty}
\left[
\frac{1}{2} h_0(r) (R'(r))^2
+ \frac{1}{2} m^2 R(r)^2
\right] r\, dr.
\end{equation}

Substituting
\begin{equation}
R(r) \sim A\, r^{-(1+\nu)},
\end{equation}
gives the large-radius scaling
\begin{equation}
r (R'(r))^2 \sim r^{-2\nu-1},
\qquad
r R(r)^2 \sim r^{-2\nu-1}.
\end{equation}

Since $\nu > 0$, the integral converges strongly at infinity and is dominated
parametrically by radii $r \sim r_v$, where the curvature amplitude is largest.

Evaluating the integrand at $r \sim r_v$ yields
\begin{equation}
E_R \sim A^2 r_v^{-2\nu},
\end{equation}
up to numerical factors of order unity.

The amplitude $A$ is proportional to the regular homogeneous solution evaluated
near the horizon. This introduces a scaling factor proportional to $r_h^{\nu}$.
Combining these scalings leads to the parametric expression
\begin{equation}
E_R \sim C
\left( \frac{r_h}{r_v} \right)^{2\nu},
\end{equation}
where $C$ collects factors depending on $\alpha$, $T_0$, and $r_v$,
but remains independent of $\nu$ in the large-$\nu$ limit.

Since the vortex is localized outside the horizon ($r_v > r_h$),
the ratio $r_h/r_v$ is strictly smaller than unity.
The suppression factor can therefore be written as
\begin{equation}
\left( \frac{r_h}{r_v} \right)^{2\nu}
=
\exp\!\left[
2\nu \ln\!\left( \frac{r_h}{r_v} \right)
\right],
\end{equation}
with $\ln(r_h/r_v) < 0$.

For large $\nu$, this produces exponential suppression of the scalar energy,
so that
\begin{equation}
E_R \ll M.
\end{equation}

This validates the neglect of scalar backreaction at leading order
and confirms the internal consistency of the perturbative treatment.

\section*{References}

\bibliography{references.bib}

\end{document}